\documentclass[reprint, aps, pra, footinbib, letterpaper, superscriptaddress]{revtex4-2}

\usepackage[T1]{fontenc} 

\usepackage{amsmath,color}
\usepackage{amssymb}
\usepackage{amsfonts}
\usepackage{amsthm}
\usepackage{mathtools}

\usepackage{algorithm}
\usepackage{algpseudocode}
\usepackage{dsfont}

\usepackage{bbold}
\usepackage{chemformula}
\usepackage{siunitx}
\usepackage{pdfpages}
\usepackage{pgffor}

\DeclarePairedDelimiterX\braket[2]{\langle}{\rangle}{#1 \delimsize\vert #2}
\DeclarePairedDelimiterX\braket3[3]{\langle}{\rangle}{#1 \delimsize\vert #2 \delimsize\vert #3}

\usepackage{accents}
\newcommand{\dbtilde}[1]{\accentset{\approx}{#1}}

\newcommand{\vE}{\mathbf{E}}

\newcommand{\ve}{\mathbf{e}}

\newcommand{\vxi}{\boldsymbol{\xi}}

\newcommand{\hH}{\hat{H}}

\newcommand{\red}[1]{{\color{black} #1}}

\makeatletter
\AtBeginDocument{\let\LS@rot\@undefined}
\makeatother

\begin{document}
	
	\title{QM/MM Modeling of Vibrational Polariton Induced Energy Transfer and Chemical Dynamics}
	
	\author{Tao E. Li}%
	\email{tao.li@yale.edu}
	\affiliation{Department of Chemistry, Yale University, New Haven, Connecticut, 06520, USA}
	
	\author{Sharon Hammes-Schiffer}%
	\email{sharon.hammes-schiffer@yale.edu}
	\affiliation{Department of Chemistry, Yale University, New Haven, Connecticut, 06520, USA}

	\begin{abstract}
        Vibrational strong coupling (VSC) provides a novel means to modify chemical reactions and energy transfer pathways. To efficiently model chemical dynamics under VSC in the collective regime, herein a hybrid quantum mechanical/molecular mechanical (QM/MM) cavity molecular dynamics (CavMD) scheme is developed and applied to an experimentally studied chemical system. This approach can achieve linear scaling with respect to the number of molecules for a dilute solution under VSC by assuming that each QM solute molecule is surrounded by an independent MM solvent bath. Application of this approach to a dilute solution of \ch{Fe(CO)5} in \textit{n}-dodecane under VSC demonstrates polariton dephasing to the dark modes and polariton-enhanced molecular nonlinear absorption. These simulations predict that strongly exciting the lower polariton may provide an energy transfer pathway that  selectively excites the equatorial \ch{CO} vibrations rather than the axial CO vibrations. Moreover, these simulations also directly probe the cavity effect on the dynamics of the \ch{Fe(CO)5} Berry pseudorotation reaction for  comparison to recent two-dimensional infrared spectroscopy experiments. This theoretical approach is applicable to a wide range of other polaritonic systems and provides a tool for exploring the use of VSC for selective infrared photochemistry.
	\end{abstract}
	
	\maketitle
    
    \section{Introduction}
    
    An intriguing finding over the past decade has been the discovery and characterization of molecular vibrational polaritons. \cite{Shalabney2015,Long2015} These hybrid light-matter states were initially observed when  a large ensemble of molecules is confined in a Fabry--P\'erot cavity, an optical device composed of a pair of parallel mirrors  supporting infrared (IR) standing electromagnetic waves. When one  standing wave, also known as a cavity mode, is near resonant with one vibrational mode of the molecules, peak splitting, or Rabi splitting, may be observed in the IR spectrum. The  Rabi splitting is a hallmark of the formation of vibrational polaritons. Under this vibrational strong coupling (VSC) domain, molecular properties may be significantly modified, including ground-state chemical reaction rates \cite{Thomas2016, Thomas2019_science} and supramolecular assembly \cite{Joseph2021} under thermal conditions, as well as intermolecular vibrational energy transfer rates under external pumping of the upper polariton (UP). \cite{Xiang2020Science} Note that challenges associated with reproducing some of the experiments under thermal conditions have been reported. \cite{Imperatore2021,Wiesehan2021}
    
    In parallel with the exciting experimental progress of vibrational polaritons, the  theoretical modeling of VSC has moved forward \cite{Li2022Review,Fregoni2022,Wang2021Roadmap} but also faces significant challenges. A major challenge on the theory side is the collective nature of polaritons versus the locality of chemical modifications. \cite{Sidler2021} Although vibrational polaritons are formed among a macroscopic number (i.e., $N\sim 10^{10}$) of molecules, many calculations of polariton effects have been performed on single-molecule VSC and therefore are not directly applicable to the collective regime. In the collective regime, each individual molecule makes a negligible contribution to the polariton and predominantly contributes to the dark modes, which are inactive with respect to electromagnetic fields. Understanding how the collective polaritonic state can induce a significant modification of local molecular properties, especially the chemical reaction rates, requires methods beyond conventional chemical modeling of molecules. Many different theoretical efforts, including the study of classical or quantum mechanical model systems, \cite{Galego2019,Campos-Gonzalez-Angulo2019,LiHuo2021,Fischer2021,YangCao2021,Wang2022JPCL} quantum-electrodynamical electronic structure theory,\cite{Flick2017,Riso2022,Schafer2021,Bonini2021,Yang2021,Philbin2022} exaction factorization, \cite{Rosenzweig2022} and the multiconfigurational time-dependent Hartree method (MCTDH), \cite{Triana2020Shape} have been directed toward modeling vibrational polaritons and their chemical effects.
    
    Another appealing approach for modeling VSC is classical cavity molecular dynamics (CavMD) simulations,\cite{Li2020Water} in which a few cavity modes coupled to a large ensemble of realistic molecules described by empirical force fields are propagated classically on an electronic ground-state surface. Beyond  classical simulations, nuclear and photonic quantum effects can be included in CavMD by path integral techniques. \cite{Li2022RPMDCav} CavMD has exhibited significant advantages for describing VSC, including the scalability of modeling a large number of molecules as well as agreement with some key experiments and analytic theory. \cite{Li2021Relaxation} One major limitation of this approach, however, is the inability to model chemical reactions under VSC. The main source of difficulty is that CavMD requires not only  nuclear forces, but also  molecular dipole moments and dipole derivatives during the time propagation. Although
    dipole derivatives may be approximated as being constant near equilibrium molecular geometries, \cite{Luk2017, Li2020Water} during chemical reactions this approximation may break down, as bond breaking and formation usually lead to very large changes in the dipole derivatives. In contrast, when electronic structure theory is used to calculate dipole derivatives on the fly as second-order energy gradients, \cite{Pulay2014} the  computational cost may become overwhelmingly expensive, especially when the molecular system size is large.  Hence, for realistic modeling of ground-state chemical reactions under collective VSC, further approximations are required to reduce the computational cost. 
    
    \begin{figure*}
		\centering
		\includegraphics[width=0.8\linewidth]{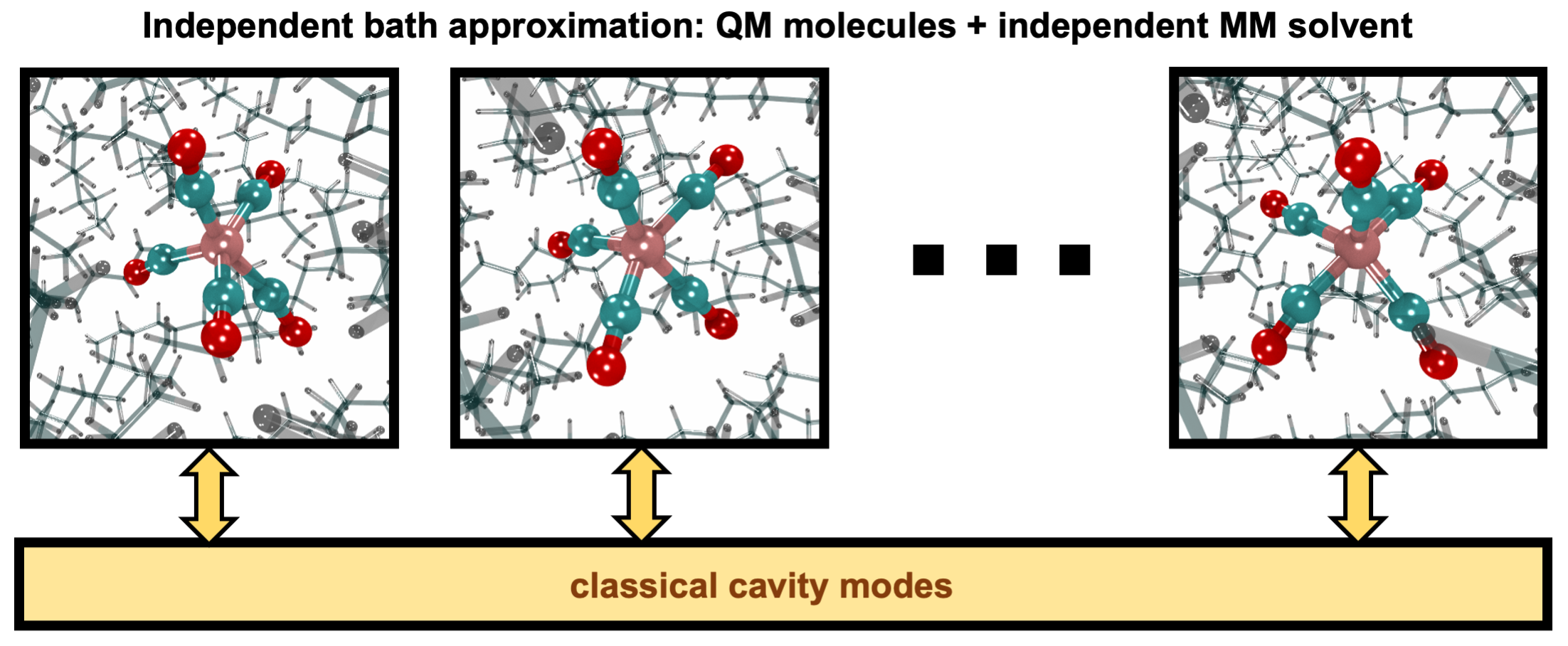}
		\caption{Illustration of the independent bath approximation for simulating collective VSC. The classical cavity modes interact with $N_b$ independent molecular systems, each of which contains a single QM molecule immersed in an  MM solvent bath. In this work, we simulate 16 \ch{Fe(CO)5} solute molecules forming VSC with a cavity mode, where each solute molecule is surrounded by 50 \textit{n}-dodecane MM molecules. At each time step, the cavity modes interact with the total dipole moment summed over all the independent molecular subsystems, and the dynamics of the molecular subsystems are  propagated in parallel. See the SI for simulation details.}
		\label{fig:setup}
	\end{figure*}
    
    Herein, we develop an efficient CavMD approach with linear scaling for dilute solutions under VSC when only the solute molecules are able to undergo chemical reactions. In this approach, the molecules are described at the hybrid quantum mechanical/molecular mechanical (QM/MM) level, \cite{Senn2009QMMMRev,Shao2007} in which the solvent molecules are described by MM with empirical force fields, and the solute molecules are described by QM with electronic structure theory. Because the solute concentration is low, we further assume no direct solute-solute interactions and invoke the independent bath approximation. As illustrated in Fig. \ref{fig:setup}, under the independent bath approximation, the cavity modes interact with $N_b$ independent molecular systems, each of which contains a single QM solute molecule surrounded by a MM solvent bath. Because the QM solute molecules are assumed to have no direct interactions with each other, their gradients can be evaluated independently and also in parallel, leading to linear scaling with respect to the number of QM molecules. The idea of the independent bath approximation was introduced in VSC by del Pino \textit{et al} \cite{Pino2015} in a fully quantum mechanical model study. The multi-scale molecular dynamics scheme for exciton-polaritons developed by Luk \textit{et al} \cite{Luk2017,Groenhof2019,Tichauer2021} also assumed the independent bath approximation.
    
    We apply this QM/MM CavMD approach to a dilute \ch{Fe(CO)5} solution in \textit{n}-dodecane under VSC. The choice of this molecular system was inspired by a recent two-dimensional infrared (2D-IR) spectroscopy experiment performed by Chen \textit{et al}. \cite{Chen2022} \red{Among many  VSC chemical reaction experiments,\cite{Thomas2016, Thomas2019_science,Nagarajan2021} this 2D-IR experiment reported the simplest reaction under VSC, namely the Berry pseudorotation reaction. \cite{Portius2019,Cahoon2008} In addition, this} experiment probed the VSC effects on the vibrational energy exchange dynamics between equatorial and axial vibrational modes. After photoexcitation of the polaritons and initial relaxation, the intramolecular vibrational energy redistribution (IVR) between the equatorial and axial vibrational modes was found to be enhanced, whereas pseudorotation was suppressed, compared to the same molecular system outside the cavity. \cite{Chen2022} These intriguing VSC effects suggest that polariton relaxation may produce a different initial distribution of the dark modes than the distribution of vibrational modes  outside the cavity.\cite{Chen2022} To better understand this experiment, which focused on the dynamics after the initial relaxation of the UP (i.e., after $\sim 2$  ps), we use the QM/MM CavMD approach to study the short-time polariton relaxation dynamics (within $\sim$ 2 ps) and investigate the resulting dark-mode distribution. 
    Our simulations show that the QM/MM CavMD approach not only reveals polariton dephasing to the dark modes as well as polariton-enhanced molecular nonlinear absorption,\cite{Li2020Nonlinear} but also predicts intriguing polariton-induced  vibrational energy transfer from the axial to  the equatorial CO vibrations in \ch{Fe(CO)5}. Moreover, this approach  can also directly simulate the cavity effect on the \ch{Fe(CO)5} Berry pseudorotation reaction \cite{Portius2019,Cahoon2008} dynamics. \red {Note that the previously developed force field CavMD approach cannot describe such reaction dynamics or the breaking and forming of chemical bonds, a distinct advantage of the QM/MM CavMD approach.} To the best of our knowledge, this study represents the first  on-the-fly first principles simulation of VSC chemical dynamics in the collective regime.

    \section{Theory and Methods}
    \label{sec:theory}
	
	\subsection{The CavMD scheme}
	
	Under VSC, because the electronic motion is much faster than the dynamics of the nuclei and IR cavity modes, the coupled nuclear-photonic dynamics can be propagated on an electronic ground-state surface. With this cavity Born--Oppenheimer approximation, \cite{Flick2017cBO} the light-matter Hamiltonian for CavMD  is defined as \cite{Li2020Water}
	\begin{subequations}\label{eq:H}
		\begin{equation}
		\begin{aligned}
		\hH_\text{QED}^{\text{G}} = \ & \hH_{\text{M}}^{\text{G}} 
		+ \hH_{\text{F}}^{\text{G}} ,
		\end{aligned}
		\end{equation}
		where $\hH_{\text{M}}^{\text{G}}$ is the conventional ground-state molecular (kinetic + potential) Hamiltonian outside a cavity, and $\hH_{\text{F}}^{\text{G}}$ denotes the field-related Hamiltonian under the long wave approximation:
		\begin{equation}\label{eq:H_QM}
		\begin{aligned}
		\hH_{\text{F}}^{\text{G}} & = \sum_{k,\lambda}
		\frac{\hat{\widetilde{p}}_{k, \lambda}^2}{2 m_{k, \lambda}} +
		\frac{1}{2} m_{k,\lambda}\omega_{k,\lambda}^2 \Bigg (
		\hat{\widetilde{q}}_{k,\lambda}  +  \frac{\varepsilon_{k,\lambda} }{m_{k,\lambda} \omega_{k,\lambda}^2}
		\hat{d}_{\text{g}, \lambda}
		\Bigg )^2 .
		\end{aligned}
		\end{equation}
		Here, $\hat{\widetilde{p}}_{k, \lambda}$, $\hat{\widetilde{q}}_{k, \lambda}$, $\omega_{k,\lambda}$, and $m_{k, \lambda}$ denote the momentum operator, position operator, frequency, and auxiliary mass for the cavity photon mode defined by a wave vector $\mathbf{k}$ and polarization direction $\vxi_\lambda$ with $\mathbf{k}\cdot \vxi_\lambda = 0$. By construction, the cavity is placed along the $z$-direction, so $\vxi_\lambda$ can be $\ve_x$ or $\ve_y$, the unit vector along the $x$- or $y$-direction.  $\hat{d}_{\text{g}, \lambda}$ denotes the electronic ground-state dipole operator for the entire molecular system projected along the direction of $\vxi_\lambda$. The quantity $\varepsilon_{k,\lambda}\equiv \sqrt{m_{k,\lambda}\omega_{k,\lambda}^2/\Omega\epsilon_0}$	characterizes the coupling strength between each cavity photon and individual molecule, where $\Omega$ denotes the cavity mode volume and $\epsilon_0$ denotes the vacuum permittivity. 
	\end{subequations}
	
	In classical CavMD, all quantum operators in Eq. \eqref{eq:H} are mapped to classical variables. Moreover, in order to use a relatively small molecular system to simulate VSC, which may involve a macroscopic number of molecules, we also assume that the entire molecular system can be represented by $N_{\text{cell}}$ identical simulation cells, \cite{Li2020Water} i.e., the total molecular dipole moment is represented by  $d_{\text{g},\lambda} = N_{\text{cell}} d_{\text{g},\lambda}^{\text{sub}}$, where $d_{\text{g},\lambda}^{\text{sub}}$ denotes the molecular dipole moment in a simulation cell. By also denoting $\dbtilde{q}_{k,\lambda} = \widetilde{q}_{k,\lambda} / \sqrt{N_{\text{cell}}}$ and an effective light-matter coupling strength
	\begin{equation}\label{eq:coupling}
	\widetilde{\varepsilon}_{k,\lambda} \equiv \sqrt{N_{\text{cell}}} \varepsilon_{k,\lambda} = \sqrt{\frac{N_{\text{cell}} m_{k,\lambda}\omega_{k,\lambda}^2}{\Omega\epsilon_0}},
	\end{equation}
	we obtain  classical equations of motion for the coupled photon-nuclear system:
	\begin{subequations}\label{eq:EOM}
	    \begin{align}
		M_{\alpha}\ddot{\mathbf{R}}_{\alpha} &= \mathbf{F}_{\alpha}^{(0)}  + \mathbf{F}_{\alpha}^{\text{cav}} 
		\\
		\label{eq:EOM-2}
		m_{k,\lambda}\ddot{\dbtilde{q}}_{k,\lambda} &= - m_{k,\lambda}\omega_{k,\lambda}^2 \dbtilde{q}_{k,\lambda}
		-\widetilde{\varepsilon}_{k,\lambda} d_{\text{g},\lambda}^{\text{sub}} + Q_{k,\lambda}\vE^{\text{ext}}(t)
	\end{align}
	\end{subequations}
	Here, the subscript $\alpha$ indexes different nuclei, and $\mathbf{F}_{\alpha}^{(0)}$ denotes the  force on each nucleus outside a cavity. Moreover,
	\begin{equation}\label{eq:F_cav}
	    \mathbf{F}_{\alpha}^{\text{cav}} = - 
	\sum_{k,\lambda}
	\left (
	\widetilde{\varepsilon}_{k,\lambda} \dbtilde{q}_{k,\lambda}
	+ \allowbreak \frac{\widetilde{\varepsilon}_{k,\lambda}^2}{m_{k,\lambda} \omega_{k,\lambda}^2} d_{\text{g},\lambda}^{\text{sub}}
	\right) 
	\frac{\partial d_{\text{g},\lambda}^{\text{sub}} }{\partial \mathbf{R}_{\alpha}}
	\end{equation}
	denotes the cavity force on each nucleus. In Eq. \eqref{eq:EOM-2}, the effective charge of the cavity modes,	$Q_{k,\lambda}$, is a phenomenological quantity that describes the coupling  coefficient between the cavity mode and the time-dependent external field $\vE^{\text{ext}}(t)$. The $Q_{k,\lambda}\vE^{\text{ext}}(t)$ term was not defined in the Hamiltonian in Eq. \eqref{eq:H} and is introduced here to phenomenologically describe the external pumping of the cavity modes. \red{From a physical perspective, this term represents an external time-dependent charge current in the normal-mode representation of  Maxwell's equations.\cite{Cohen-Tannoudji1997} The $Q_{k,\lambda}\vE^{\text{ext}}(t)$ term  also corresponds to the input mode in the input-output theory of quantum optics.\cite{Gardiner2004,Ribeiro2018Nonlinear} Although the value of $Q_{k,\lambda}$ should be an intrinsic property of an optical cavity, throughout this manuscript, we set $Q_{k,\lambda}$  as 0.1 a.u. for simplicity.} Although previous CavMD simulations usually excite polaritons by pumping the molecular subsystem, \cite{Li2020Nonlinear,Li2021Relaxation} here the polaritons are excited by pumping the cavity mode. These two approaches have been found to yield similar results for the polariton relaxation dynamics, \cite{Li2021Solute} as long as the coherent energy exchange between the molecular bright state and the cavity mode, as quantified by the Rabi splitting, is much faster than either the cavity loss or the molecular dissipation.
	
	In order to propagate the time-dependent CavMD dynamics governed by Eq. \eqref{eq:EOM}, we need to evaluate three key quantities: $\mathbf{F}_{\alpha}^{(0)}$, $d_{\text{g},\lambda}^{\text{sub}}$, and $\partial d_{\text{g},\lambda}^{\text{sub}}/\partial \mathbf{R}_{\alpha}$. 
	Previous CavMD studies used empirical MM force fields, \cite{Li2020Water} for which each atom is assigned a fixed point charge $Q_{\alpha}$. At this level of theory, $\mathbf{F}_{\alpha}^{(0)}$ can be easily evaluated by standard MD packages, and the dipole moment and dipole derivatives can also be calculated in a straightforward manner:
	\begin{subequations}\label{eq:dipole_evaluation_MM}
	    \begin{align}
	        d_{\text{g},\lambda}^{\text{sub}} &= \sum_{\alpha} Q_{\alpha} \mathbf{R}_{\alpha} \cdot \vxi_{\lambda}, \\
	        \frac{\partial d_{\text{g},\lambda}^{\text{sub}} }{\partial R_{\alpha i}} &= Q_{\alpha} \ve_i \cdot \vxi_{\lambda}, 
	    \end{align}
	\end{subequations}
	where $i = x, y, z$ indexes the Cartesian components of the nuclear coordinates and $\mathbf{R}_{\alpha} = (R_{\alpha x}, R_{\alpha y}, R_{\alpha z})$. Because the computational cost of evaluating Eqs. \eqref{eq:F_cav} and \eqref{eq:dipole_evaluation_MM} is marginal compared with calculating $\mathbf{F}_{\alpha}^{(0)}$, the computational cost of CavMD with MM force fields is similar to the cost of conventional MD outside the cavity.
	
	\subsection{The independent bath approximation}
	
	As illustrated in Fig. \ref{fig:setup}, when a dilute solution under VSC is considered, we invoke the independent bath approximation. Under this approximation, the classical cavity modes interact with $N_b$ independent molecular systems, each of which is treated with the standard QM/MM scheme, i.e., a single QM molecule is immersed in a bath of solvent molecules described by an MM force field, and the interaction between the QM and MM systems can be treated with either mechanical embedding \cite{Vreven2006ONIOM} or electrostatic embedding. \cite{Senn2009QMMMRev} Within this QM/MM treatment, the nuclear forces outside the cavity, $\mathbf{F}_{\alpha}^{(0)}$, are standard outputs of  QM/MM packages. For mechanical embedding treatments, the  total dipole moment and dipole derivatives may be expressed as 
	\begin{subequations}\label{eq:dipole_evaluation_QMMM}
	    \begin{align}
	    \label{eq:dipole_evaluation_QMMM-a}
	        d_{\text{g},\lambda}^{\text{sub}} &= \sum_{\beta = 1}^{N_b} \left( d_{\text{g},\lambda}^{\beta, \text{QM}} + d_{\text{g},\lambda}^{\beta, \text{MM}} \right ), \\
	        \frac{\partial d_{\text{g},\lambda}^{\text{sub}} }{\partial R_{\alpha i}^{\beta}} &=
	        \label{eq:dipole_evaluation_QMMM-b}
	        \begin{cases}
            \frac{\partial d_{\text{g},\lambda}^{\beta, \text{QM}} }{\partial R_{\alpha i}^{\beta}}  &\text{if  $\alpha \in$ QM},\\
            Q_{\alpha}^{\beta} \ve_i \cdot \vxi_{\lambda}  &\text{if  $\alpha \in$ MM}.
            \end{cases}
	    \end{align}
	\end{subequations}
	In Eq. \eqref{eq:dipole_evaluation_QMMM-a}, $\beta = 1, 2, \cdots, N_b$ indexes different independent molecular systems, and in each independent molecular system the total dipole moment is simply the sum of the QM dipole moment $d_{\text{g},\lambda}^{\beta, \text{QM}}$ and the MM dipole moment $d_{\text{g},\lambda}^{\beta, \text{MM}}$, the latter of which is calculated by Eq. \eqref{eq:dipole_evaluation_MM}a. In Eq. \eqref{eq:dipole_evaluation_QMMM-b}, for the QM dipole derivatives $\partial d_{\text{g},\lambda}^{\beta, \text{QM}} /\partial R_{\alpha i}^{\beta}$, the off-diagonal components such as $\partial d_{\text{g},x}^{\beta, \text{QM}} /\partial R_{\alpha z}^{\beta}$ (when $\lambda = x$ and $i = z$) are usually non-zero. This behavior is very different from the MM dipole derivatives $Q_{\alpha}^{\beta} \ve_i \cdot \vxi_{\lambda}$, which is always zero when $i \neq \lambda$. Note that Eq. \eqref{eq:dipole_evaluation_QMMM-b} is exact for mechanical embedding but would need to be modified for electrostatic embedding QM/MM implementations.
	
	Although the solvent molecules are treated with MM force fields, Eq. \eqref{eq:dipole_evaluation_QMMM} guarantees that both the solute and the solvent molecules interact with the cavity modes. 
	If the solvent molecules do not form VSC due to a lack of vibrational modes near the frequencies of the cavity modes, we may further replace the MM contributions in Eq. \eqref{eq:dipole_evaluation_QMMM} with zeros. 
	
	With the independent bath approximation, the computational cost scales linearly with the number of QM molecules. Moreover, because the gradients in each local bath can be computed in parallel, first-principles CavMD can  be used to simulate collective VSC with an acceptable computational cost. For our simulations, the QM/MM CavMD approach is implemented by interfacing the molecular dynamics package I-PI \cite{Kapil2019} with the stand-alone QM/MM gradients in Q-Chem. \cite{Woodcock2007QChemQMMM,Epifanovsky2021}

	\subsection{Simulation Details}
	\label{sec:simulation_details}
    As an illustrative example of QM/MM CavMD, we simulate the VSC effects on a dilute solution of \ch{Fe(CO)5} in \textit{n}-dodecane solvent at 300 K. For this system, as depicted in Fig. \ref{fig:setup}, VSC is formed between a cavity mode (with both $x$- and $y$-polarizations) and the \ch{CO} vibrations of the  \ch{Fe(CO)5} molecules. Under the independent bath approximation, we include 16 QM \ch{Fe(CO)5} molecules described by density functional theory (DFT) with the BP86 \cite{Becke1988} functional  and def2-mSVP \cite{Grimme2015def2-msvp} basis set. Each \ch{Fe(CO)5} molecule is surrounded by 50 MM \textit{n}-dodecane molecules described by the OPLS-AA \cite{Jorgensen1996OPLSAA} force field. \red{A two-layer ONIOM model with} mechanical embedding \cite{Vreven2006ONIOM} is used to characterize the interaction between the QM and MM regions. No cavity loss is included in these simulations. Note that including only 16 \ch{Fe(CO)5} molecules in the simulation is expected to yield similar short-time polariton relaxation dynamics as would be obtained for a large number of molecules under VSC, as long as the Rabi splitting is the same for the two cases. This convergent behavior was shown previously with MM force fields, \cite{Li2021Solute} and the current QM/MM results are expected to behave similarly. See the SI for more simulation details.
    
	\section{Results and Discussion}

	\subsection{Polariton spectrum}
    
    \begin{figure*}
		\centering
		\includegraphics[width=0.8\linewidth]{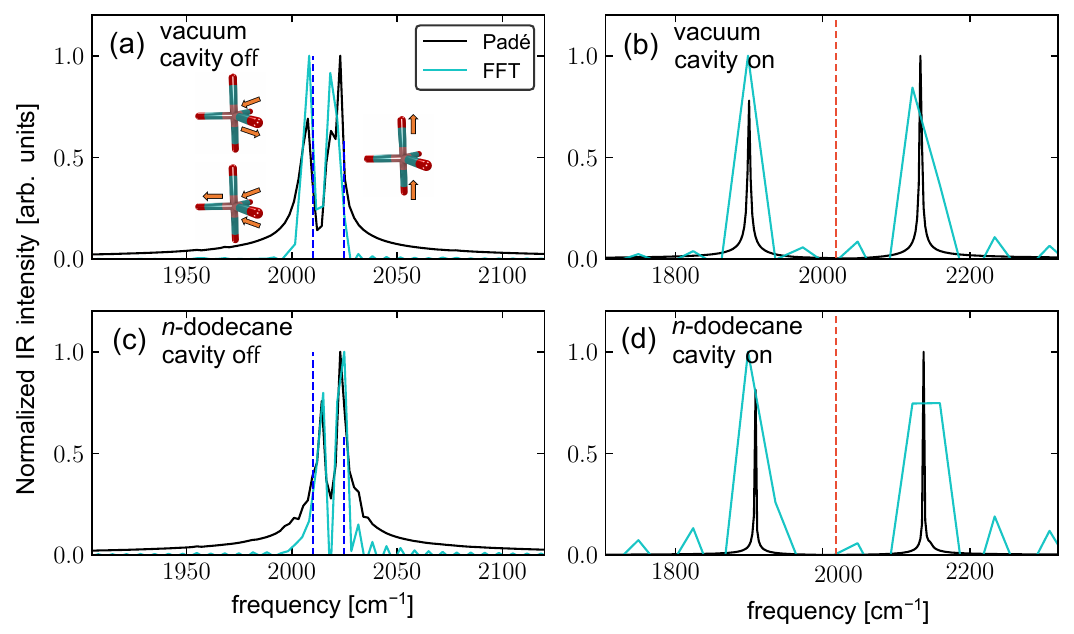}
		\caption{(a) IR spectrum of a single \ch{Fe(CO)5} in vacuum outside the cavity evaluated by Fourier transforming the dipole autocorrelation function.  The Pad\'e  transform (solid black line) of the autocorrelation function exhibits much better frequency resolution than the fast Fourier transform (solid cyan line). The vertical dashed lines denote the IR spectrum evaluated by a Hessian calculation in vacuum. \red{The inset shows the axial and equatorial modes, which are linear combinations of CO stretch modes.} (b) The corresponding polariton spectrum when 16 independent, noninteracting  \ch{Fe(CO)5} molecules in vacuum are coupled to the cavity mode at $\omega_c=$ 2018 cm$^{-1}$ (vertical dashed red line). (c,d) The same plots as the upper panel except that the solvent environment is explicitly modeled by MM \textit{n}-dodecane molecules.}
		\label{fig:spectrum}
	\end{figure*}
	
	Before presenting the polariton spectrum, we focus on the IR spectrum of \ch{Fe(CO)5} in vacuum. As shown in Fig. \ref{fig:spectrum}a, for a single \ch{Fe(CO)5} in vacuum at 300 K, the IR spectrum of this molecule is calculated by evaluating the Fourier transform of the dipole autocorrelation function from a 10 ps NVE (constant number of particles, volume, and energy) trajectory. Because the trajectory is relatively short, the direct Fourier transform (solid cyan line) has relatively low resolution. A Pad\'e  approximation \cite{Bruner2016,Goings2018} of the Fourier transform (solid black line) provides higher resolution. Both simulated spectra show a two-peak feature near 2000 cm$^{-1}$, where the lower frequency peak reflects the doubly degenerated equatorial \ch{CO} vibration with $e'$ symmetry and the higher frequency peak reflects the axial \ch{CO} vibration with $a_2''$ symmetry;\cite{Portius2019} see the cartoons inserted in Fig. \ref{fig:spectrum}a for the corresponding vibrational normal modes.  The peak positions from the dipole autocorrelation function agree with the Hessian calculation of a single \ch{Fe(CO)5} at its equilibrium geometry, as indicated by the dashed vertical blue lines (at 2010 cm$^{-1}$ for the $e'$ vibration and 2025 cm$^{-1}$ for the $a_2''$ vibration). The heights of the dashed vertical  blue lines are proportional to the IR intensities from the Hessian calculation. 
	
	Next we investigate 16 randomly oriented \ch{Fe(CO)5} molecules in vacuum  coupled to the cavity mode  with the coupling strength $\widetilde{\varepsilon} = 10^{-4}$ a.u.\red{, which, according to Eq. \eqref{eq:coupling}, corresponds to a cavity volume of $\Omega = N_{\text{cell}} m_c \omega_c^2 /\widetilde{\varepsilon}^2\epsilon_0 \sim 10\ N_{\text{cell}}$ nm$^3$}.  Fig. \ref{fig:spectrum}b plots the polariton spectrum by Fourier transforming the photon autocorrelation function calculated from a 1 ps CavMD simulation. The cavity mode frequency of $\omega_c = 2018$ cm$^{-1}$ is indicated by the vertical dashed red line. Although such a short trajectory cannot provide sufficient frequency resolution when a direct Fourier transform (solid cyan line) is applied, the Pad\'e  approximation predicts the formation of a LP and an UP, with frequencies at $\omega_{\text{LP}} = 1900$ cm$^{-1}$ and $\omega_{\text{UP}} = 2133$ cm$^{-1}$, respectively. \red{The relative heights of the UP and LP predicted by FFT and the P\'ade approximation are slightly different, most likely due to the fitting error introduced by the P\'ade approximation.} The middle polariton (MP) formed by the two vibrational modes coupled to one cavity mode, \cite{Xiang2020Science} which is usually very weak, cannot be identified from this short trajectory.
	
	The corresponding spectra obtained when the \textit{n}-dodecane solvent molecules are included in the simulation are plotted in Figs. \ref{fig:spectrum}c,d. The spectra computed both inside and outside the cavity in the presence of \textit{n}-dodecane are changed only slightly compared to the analogous vacuum spectra. Specifically, inside the cavity, the LP and UP frequencies become $\omega_{\text{LP}} = 1909$ cm$^{-1}$ and $\omega_{\text{UP}} = 2137$ cm$^{-1}$, which gives a Rabi splitting of $\Omega_R = 228$ cm$^{-1}$, compared to $\Omega_R = 233$ cm$^{-1}$ in vacuum. The relatively small impact of solvent on these IR spectra is understandable, given that \textit{n}-dodecane is a non-polar solvent.

	\subsection{Polariton relaxation dynamics}

	\begin{figure*}
		\centering
		\includegraphics[width=0.8\linewidth]{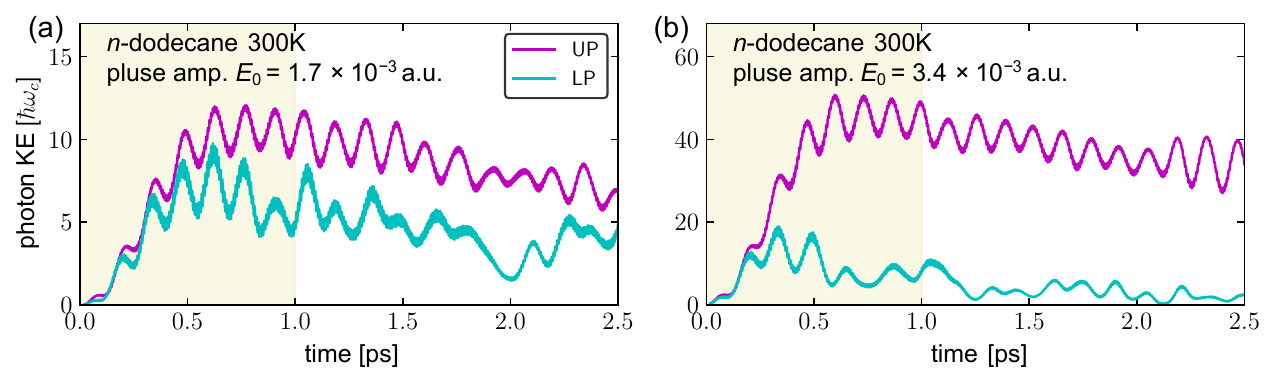}
		\caption{(a) Kinetic energy dynamics of cavity photons under Gaussian pumping of the UP (magenta curve) or LP (cyan curve). The shadowed yellow window represents the time domain during which the Gaussian pulse is turned on. Because the cavity loss rate is assumed to be zero, the decay of the photonic energy indicates  polariton relaxation to the dark modes. (b) The same plot as part (a) except that the pulse amplitude is doubled. Although the UP dynamics remains similar as in part (a), the LP dissipation is greatly accelerated, implicating the LP-enhanced molecular nonlinear absorption mechanism. Each data point here has been smoothed using adjacent averaging over 40-fs windows to remove the fast oscillations in the kinetic energy.
		}
		\label{fig:photon_dynamics_300K}
	\end{figure*}

	After obtaining the polariton spectrum, we explore the nonequilibrium polaritonic dynamics for the dilute solution of \ch{Fe(CO)5} in \textit{n}-dodecane. The nonequilibrium polaritonic dynamics is initialized by pumping the cavity mode with a Gaussian pulse:
	\begin{equation}\label{eq:E_gaussian}
        \vE(t) = \ve_x E_0\sin(\omega t)\exp\left [-2\ln(2)(t/\tau)^2\right ].
    \end{equation}
    This pulse was applied during the time window $0 < t < 1$ ps with a width of $\tau = 0.5$ ps.
	
	First we simulate the nonequilibrium relaxation dynamics when the amplitude of the Gaussian pulse is set as $E_0 = 1.7\times 10^{-3}$ a.u. Fig. \ref{fig:photon_dynamics_300K}a plots the photon kinetic energy \red{($\sum_{k,\lambda} p_{k,\lambda}^2/2m_{k,\lambda}$)} dynamics when either the UP (magenta line) or the LP (cyan line) is pumped. At this pumping amplitude, both the UP and LP excitations demonstrate similar energy gain and relaxation dynamics. The oscillations of the photonic signals have a period of $\sim$ 0.15 ps, which reflects the coherent energy transfer between the cavity photon and molecular bright state, in agreement with the Rabi splitting ($\Omega_R = 228$ cm$^{-1}$ = 0.146 ps). At $t = 2.5$ ps, the photonic signals are dissipated to half the initial energy.
	Because cavity loss is assumed to be zero and vibrational energy relaxation to the ground state is much slower than a few ps, the photonic energy relaxation on a ps time scale after the Gaussian pulse implies polaritonic energy transfer to the dark modes due to a dephasing mechanism. \red{This dephasing behavior agrees with the previous force field CavMD simulations of liquid \ch{CO2} under VSC (see Fig. 3 therein),\cite{Li2020Nonlinear} cross-validating both the QM/MM and force field CavMD approaches, although the previous CavMD simulations did not invoke the independent bath approximation.}
	
	\red{In these simulations, the molecules are rotating in a manner that changes their orientations and thus influences their coupling to the cavity mode, thereby altering the relative weightings of the molecules in the polaritonic states along the trajectory. In  the absence of solvent molecules, the UP dephasing dynamics are found to be greatly suppressed because the solvent phonon modes facilitate relaxation from the UP to the dark modes \cite{Pino2015} (see Figure S1 in the SI). In contrast, the LP dephasing dynamics are similar in the gas phase and solution for this system, suggesting that the dominant relaxation pathways from the LP state are influenced mainly by vibrational anharmonicity of the solute. Pumping the UP state allows relaxation to the dark modes, but subsequent relaxation to the LP state is entropically unfavorable because the number of dark modes is much larger than the single LP state. Pumping the LP state allows relaxation to the dark modes when the Rabi splitting is less than or approximately equal to the thermal energy,\cite{Li2021Relaxation} as is the case in the present study. Polariton relaxation dynamics when the Rabi splitting is much larger than room temperature or in a polar solvent environment will be interesting to explore.} 
	
	Qualitatively different behavior is observed when the pulse amplitude is increased by a factor of two. Fig. \ref{fig:photon_dynamics_300K}b plots the corresponding photon kinetic energy dynamics after pumping the LP or the UP with this stronger pulse.  The UP dynamics remains similar to the dynamics observed for the weak pumping case, except that the maximal value is increased by a factor of four and the decay is slightly slower. In contrast, the LP decay dynamics becomes much faster than the dynamics in Fig. \ref{fig:photon_dynamics_300K}a.  The fast LP decay under strong pumping has also been observed in force field CavMD simulations of liquid \ch{CO2} under VSC \cite{Li2020Nonlinear} and has been shown to be caused by the LP-enhanced molecular nonlinear absorption mechanism.  \cite{Xiang2019State,Xiang2021JCP,Ribeiro2020,Li2020Nonlinear}
	
	This nonlinear mechanism can be understood in both quantum and classical mechanical pictures. From the quantum mechanical perspective, because molecular vibrations are anharmonic, it is possible to prepare a cavity setup such that twice the LP frequency roughly matches the $0 \rightarrow 2$ vibrational transition. In this case, under strong excitation, the energy of two LP quanta can be directly transferred to the second excited vibrational state of individual molecules. Thus, this nonlinear mechanism  provides an additional decay channel for the LP, leading to a shorter LP lifetime under strong excitation. From the classical perspective, LP-enhanced molecular nonlinear absorption can be understood as a ``self-catalysis'' mechanism: once an individual molecular vibration gains energy during the LP dephasing, its instantaneous vibrational frequency becomes red-shifted due to anharmonicity, causing a larger spectral overlap with the LP and leading to an even faster energy loss of the LP.

	\subsection{Polariton induced energy transfer}
	
	\begin{figure}
		\centering
		\includegraphics[width=1.0\linewidth]{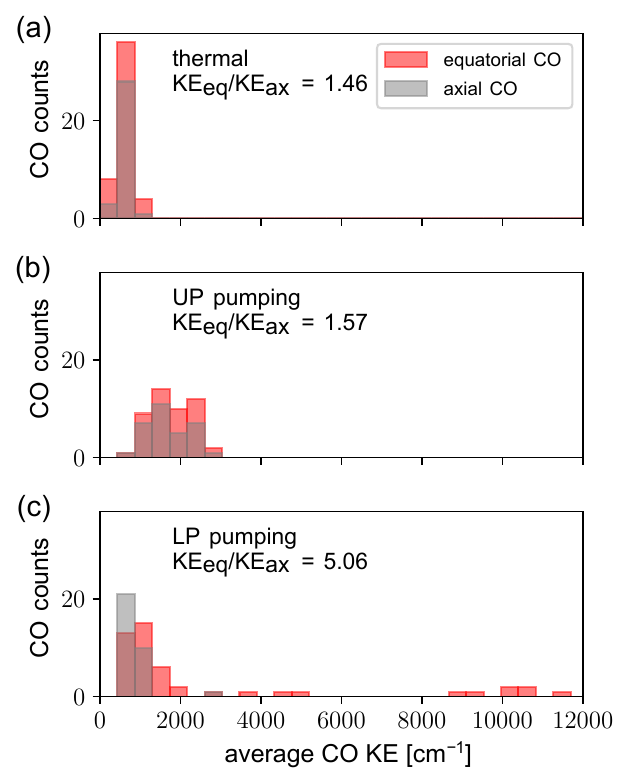}
		\caption{Transient kinetic energy distribution of the \ch{CO} bonds  under three conditions: (a) thermal conditions, (b) UP pumping, and (c) LP pumping. The red bars denote the equatorial \ch{CO} bonds, and the gray bras denote the  axial \ch{CO} bonds. The simulation conditions under external pumping are the same as for Fig. \ref{fig:photon_dynamics_300K}b. The transient kinetic energy of each \ch{CO} group is calculated by averaging over a time window $1.25 < t < 1.75$ ps to minimize the thermal noise. Part (c) shows that pumping the LP can selectively excite the equatorial bonds.}
		\label{fig:VCO_dist}
	\end{figure}
	
	After studying the polaritonic energy dynamics following Gaussian pumping, we investigate the \ch{CO} vibrational energy distribution among the \ch{Fe(CO)5}  molecules during the polariton relaxation. As a control, Fig. \ref{fig:VCO_dist}a plots the kinetic energy statistics of each local \ch{CO} group in all \ch{Fe(CO)5} molecules in the absence of polariton pumping. The equatorial and axial \ch{CO} groups are indicated by the red and gray bars, respectively.
	In order to reduce the thermal noise, the kinetic energy of each \ch{CO} group is calculated by averaging the kinetic energies of the \ch{C} plus \ch{O} atoms during a $0.5$ ps time window. As expected, under thermal conditions, both the equatorial and axial \ch{CO} groups exhibit a Maxwell--Boltzmann distribution centered at $\sim 3k_\text{B} T = 626$ cm$^{-1}$, which is equal to the thermal kinetic energy of each \ch{CO} group at 300 K. The kinetic energy ratio between all equatorial and all axial \ch{CO} groups is $1.46$, in agreement with the fact that each \ch{Fe(CO)5} molecule contains three equatorial \ch{CO}  and two axial \ch{CO} bonds.
	
	Fig. \ref{fig:VCO_dist}b plots the transient kinetic energy statistics of each local \ch{CO} group  after  strong UP excitation, under the same simulation conditions as Fig. \ref{fig:photon_dynamics_300K}b.  The transient kinetic energy of each \ch{CO} group is calculated by averaging the kinetic energy of the \ch{CO} group during the time window $1.25 < t < 1.75$ ps. Compared with the thermal distribution in Fig. \ref{fig:VCO_dist}a, the UP pumping causes the kinetic energy distribution to be blue-shifted and broadened. The broadened distribution is centered around $2000$ cm$^{-1}$. In Fig. \ref{fig:VCO_dist}b, the kinetic energy ratio between all equatorial and all axial \ch{CO} groups is $1.57$. This ratio implies that, compared with the thermal case, exciting the UP does not cause a significantly different vibrational energy distribution between equatorial and axial bonds. This ratio, however, does not indicate that exciting the UP would not cause vibrational energy transfer between the equatorial and axial bonds. The weights of the various vibrations in the polariton state, as quantified by the Hopfield coefficients\cite{Xiang2020Science}, can be very different from the population of each vibration after the polariton relaxation. This difference can be used to quantify the magnitude of polariton-induced vibrational energy transfer.  According to the simple three-state model in the SI, the ratio of Hopfield coefficients for the equatorial and axial vibrations is 1.33. The larger ratio in Fig. \ref{fig:VCO_dist}b might indicate an 18\% (i.e., $1.57/1.33-1$) enhancement in vibrational energy transfer from the axial to the equatorial vibration. In order to obtain a more accurate estimation of the polariton effect on vibrational energy transfer, more extensive simulations are needed to reduce the thermal noise.

	 Analogous to the study of the UP excitation, Fig. \ref{fig:VCO_dist}c plots the corresponding transient vibrational energy statistics of the \ch{CO} groups after the LP excitation, under the same simulation conditions as Fig. \ref{fig:photon_dynamics_300K}b.  In contrast to Fig. \ref{fig:VCO_dist}b, we observe a very different vibrational energy distribution between the equatorial (red bars) and axial (gray bars) \ch{CO} groups. The kinetic energy ratio between all equatorial and all axial \ch{CO} groups now becomes 5.06, which  surpasses both the thermal and UP pumping cases by more than a factor of two. This value is also much larger than the Hopfield coefficients ratio, which is 2.2 as calculated in the SI, showing significant  vibrational energy transfer from the axial to the equatorial vibration caused by the LP pumping. 
	 
	 Moreover, the equatorial \ch{CO} groups now dominate the high-energy tail of the distribution, showing that the equatorial \ch{CO} groups are more likely to be strongly excited. This selective excitation of the equatorial groups, which is reminiscent of a  previous finding of selective excitation of the solute molecules after strong solvent LP pumping, \cite{Li2021Solute} is likely to result from the LP-enhanced molecular nonlinear absorption mechanism. Compared with the axial \ch{CO} group, the slightly lower vibrational frequency of the equatorial \ch{CO} group appears to cause a stronger interaction with the LP, which in turn leads to more vibrational energy redistribution toward the high-energy tail. 
	 
	 \subsection{Chemical dynamics under polariton relaxation}

\begin{figure}
		\centering
		\includegraphics[width=1.0\linewidth]{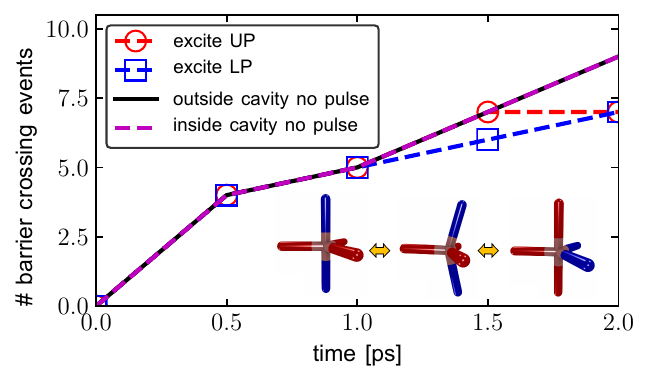}
		\caption{Number of barrier crossing events during \ch{Fe(CO)5} pseudorotation as a function of simulation time. Four cases are compared: outside (solid black) and inside (dashed magenta) the cavity under thermal conditions, and exciting the UP (dashed red) or the LP (dashed blue) inside the cavity. The initial coordinates and velocities for all molecules in the 16 independent molecular systems are the same for the four cases.}
		\label{fig:chem_dynamics}
	\end{figure}

	 Beyond the energy transfer analysis, the real advantage of QM/MM CavMD is the ability to simulate chemical reaction dynamics under VSC. At room temperature, \ch{Fe(CO)5} can undergo geometry  isomerization by exchanging the two axial \ch{CO} ligands with two of the equatorial CO ligands, known as the Berry pseudorotation reaction, \cite{Portius2019,Cahoon2008} as shown in the inset of Fig. \ref{fig:chem_dynamics}. Although it does not involve bond breaking and formation, this isomerization process can be regarded as a simple chemical "reaction", which involves a low-energy transition state and occurs on a ps time scale. \cite{Cahoon2008} Such a fast geometry isomerization can be tracked on the fly by 2D-IR spectroscopy \cite{Cahoon2008} and can also be observed during our QM/MM simulations. 
	 
	 Fig. \ref{fig:chem_dynamics} plots the number of barrier crossing events during the pseudorotation as a function of time for the simulation outside (solid black) and inside (dashed magenta) the cavity in the absence of external pumping.
	 Without pumping, the reaction dynamics inside and outside the cavity are the same, showing that VSC under thermal conditions does not significantly modify the Berry pseudorotation reaction.   Under external pumping of the UP (dashed red) or the LP (dashed blue), the Berry pseudorotation reaction seems to slow down slightly. 
	 Because the initial states (i.e., the coordinates and velocities) of the molecules for all the simulations in Fig. \ref{fig:chem_dynamics} are exactly the same and the simulations are performed deterministically under an NVE ensemble, such a difference indicates that the IR photochemical dynamics can be tuned by pumping  polaritons.  This cavity suppression of the pseudorotation is consistent with observations from the recent 2D-IR experiments. \cite{Chen2022} Due to the relatively small number of independent molecular systems studied, however, these results are not statistically reliable for comparison to experiments and should be viewed only qualitatively. One possible explanation of the polariton effect on pseudorotation is a solvent effect, in which polariton relaxation to the dark modes excites the low-frequency phonon modes of the nearby solvent molecules, which in turn hinder the pseudorotation dynamics. Another possible explanation is that excitation of other molecular modes of the \ch{Fe(CO)5} hinders the pseudorotation dynamics. To investigate these potential mechanisms, more trajectories and additional analyses are required. 
	 
	 \red{Lastly, although LP pumping can create highly excited equatorial vibrations, as compared to the case of UP pumping, generating these highly excited \ch{CO} vibrations does not appear to significantly modify the pseudorotation dynamics (see Figs. \ref{fig:VCO_dist} and \ref{fig:chem_dynamics}). These findings are not contradictory because pseudorotation is controlled by the rotation of the \ch{Fe-C} bonds, and excitation of the \ch{CO} vibrations is not expected to modify these \ch{Fe-C} rotations.}

	\section{Conclusion}
	
	This paper presents the QM/MM CavMD scheme for describing chemical dynamics under collective VSC and the application of this approach to a dilute solution of \ch{Fe(CO)5} in \textit{n}-dodecane under VSC, which has been studied experimentally.\cite{Chen2022} According to the simulations,  external pumping of the polariton leads to polariton dephasing to the dark modes as well as polariton-enhanced molecular nonlinear absorption.\cite{Li2020Nonlinear} Moreover, after strong pumping of the LP, the vibrational energy was found to transfer preferably to the equatorial \ch{CO} vibrations rather than the axial CO vibrations, causing the equatorial vibrational modes to become highly excited. This nontrivial energy transfer pathway highlights the complexity of vibrational energy transfer dynamics under polariton pumping. These QM/MM CavMD simulations also suggest that pumping polaritons could potentially suppress the Berry pseudorotation dynamics. 
	
	The recent 2D-IR experiments and accompanying analysis \cite{Chen2022} of \ch{Fe(CO)5} solution under VSC focused on the dynamics after the initial polariton relaxation (i.e., after $\sim$ 2 ps). Our simulations of the short-time polariton relaxation stage (i.e., within $\sim 2$ ps) provide insights into the distribution of vibrational modes following the polariton relaxation over the initial 2 ps. This distribution is expected to influence the subsequent dynamics studied experimentally. Moreover, these  simulations may assist future interpretations of the shorter timescale spectroscopic data, which is challenging. Additional simulations and further analysis may provide an explanation for the experimentally observed and computationally suggested suppression of the Berry pseudorotation dynamics. Beyond this specific system, the QM/MM CavMD approach developed in this work provides a numerical tool for exploring the possibility of using VSC to achieve selective IR photochemistry in the liquid phase.
	
	\section{Supporting information}
	
	Additional simulation details; three-state Hopfield model for calculating the Hopfield coefficients for each polariton; polariton relaxation dynamics for \ch{Fe(CO)5} in vacuum; CavMD simulations of VSC effects on single-molecule isomerization reaction.
	The source code, input files, and related tutorials are available at Github   (\url{https://github.com/TaoELi/cavity-md-ipi}).
	
	\section{Acknowledgements}
	
    This material is based upon work supported by the Air Force Office of Scientific Research under AFOSR Award No. FA9550-18-1-0134. We thank Dr. Tengteng Chen and Prof. Wei Xiong for
    useful discussions.

	
	\providecommand{\latin}[1]{#1}
	\makeatletter
	\providecommand{\doi}
	{\begingroup\let\do\@makeother\dospecials
		\catcode`\{=1 \catcode`\}=2 \doi@aux}
	\providecommand{\doi@aux}[1]{\endgroup\texttt{#1}}
	\makeatother
	\providecommand*\mcitethebibliography{\thebibliography}
	\csname @ifundefined\endcsname{endmcitethebibliography}
	{\let\endmcitethebibliography\endthebibliography}{}

	\clearpage
	\onecolumngrid
	

\section*{Supplementary material}

# Supporting Information

# QM/MM Modeling of Vibrational Polariton Induced Energy Transfer and Chemical Dynamics

Tao E. Li* and Sharon Hammes-Schiffer*

*Department of Chemistry, Yale University, New Haven, Connecticut, 06520, USA*

E-mail: tao.li@yale.edu; sharon.hammes-schiffer@yale.edu

# 1. Additional simulation details

In this work, a molecular system of 16 $Fe(CO)_5$ molecules is coupled to a cavity mode with frequency $\omega_c = 2018$ cm$^{-1}$. The cavity mode contains two ($x$ and $y$) polarization directions. The auxiliary mass of the cavity photon mode is set as $m_c = m_{k,\lambda} = 1$ a.u. The $Fe(CO)_5$ molecules are described by DFT with the BP86[S1] functional, which has been used extensively for $Fe(CO)_5$,[S2,S3] and the def2-mSVP[S4] basis set. For each individual $Fe(CO)_5$ molecule, the surrounding solvent environment is modeled by 50 $n$-dodecane molecules using the OPLS-AA[S5] force field. The QM/MM interaction is treated with a two-layer ONIOM model under mechanical embedding,[S6] which is expected to be reasonable for this relatively non-polar solvent, as implemented in Q-Chem. In the QM/MM calculations, the partial charges and Lennard–Jones parameters of $Fe(CO)_5$ are defined as follows: $Q_{Fe} = 0.0238\ |e|$, $\sigma_{Fe} = 3.11$ Å, $\epsilon_{Fe} = 0.2868$ kcal/mol; $Q_C = 0.07124\ |e|$, $\sigma_C = 3.30$ Å, $\epsilon_C = 0.6597$ kcal/mol; $Q_O = -0.0760\ |e|$, $\sigma_O = 3.09$ Å, $\epsilon_O = 0.1434$ kcal/mol. The Lennard–Jones parameters were chosen from the OPLS-AA force field, and the partial charges were calculated by the Hirshfeld population analysis[S7] from the QM calculation for a single $Fe(CO)_5$ molecule at its optimized geometry. Although the Hirshfeld partial charges of C and O atoms at equatorial and axial locations are slightly different, here, due to the existence of pseudorotation, $Q_C$ ($Q_O$) is defined as the average value of the five C (O) Hirshfeld partial charges.

Because the $n$-dodecane vibrational frequencies differ significantly from the CO vibrations in $Fe(CO)_5$, we chose to turn off the cavity-solvent interaction throughout the calculations presented in this manuscript, i.e., in Eq. (6), the MM contribution to the dipole moment and dipole derivatives is set to zero.

The initial geometry of the QM/MM system was prepared as follows. The configuration of one $Fe(CO)_5$ molecule surrounded by 50 $n$-dodecane molecules was prepared in a cubic cell of length 13.4 Å by PACKMOL.[S8] This initial configuration, with initial velocities chosen according to a Maxwell–Boltzmann distribution at 300K, was equilibrated for 20 ps under an NVT (constant number of particles, volume, and temperature) ensemble attached to a

Langevin thermostat with a friction lifetime of 100 fs. From this 20 ps NVT trajectory, 16 configurations were taken from $t = 12.5$ ps to $t = 20$ ps with an inverval of 0.5 ps. Each of these 16 configurations, which contain one $Fe(CO)_5$ molecule plus 50 $n$-dodecane molecules, was rotated randomly with PACKMOL, and the combination of these 16 configurations formed the initial state of the VSC simulations. Because the combined molecular configurations in this initial state may have a permanent dipole moment, in order to ensure the initial potential energy for each photon coordinate is thermally equilibrated, we set the initial photonic coordinates to be $\tilde{\tilde{q}}_{c,\lambda} = \sqrt{k_B T/m_c \omega_c^2} - \tilde{\varepsilon} d_{g,\lambda}^{sub}/m_c \omega_c$, so that the potential energy in Eq. (1b) for each photon coordinate was $k_B T/2$.

With the initial QM/MM molecular system configuration prepared as described above, we simulated the coupled cavity-molecular system for 1 ps to calculate the polariton spectrum. The initial velocities of all particles (photons + nuclei) were assigned according to the Maxwell–Boltzmann distribution at 300K. This simulation was propagated for 1 ps under an NVE ensemble. Note that the cavity mode equation of motion given in Eq. (3b) depends on the total dipole moment summed over all the independent molecular systems at each time step, which therefore must be propagated in parallel. When calculating the polariton spectrum, the Fourier transform of the autocorrelation function of the photonic coordinates is calculated:

$$C(\omega) = \frac{\omega^2}{2\pi} \int_{-\infty}^{+\infty} dt \; e^{-i\omega t} \left\langle \sum_{\lambda=x,y} \tilde{\tilde{q}}_{c,\lambda}(0) \tilde{\tilde{q}}_{c,\lambda}(t) \right\rangle \tag{S1}$$

With the same initial QM/MM molecular configurations as used to calculate the polariton spectrum, we propagated a nonequilibrium simulation of the coupled cavity-molecular system under a Gaussian pulse excitation for 2.5 ps. The Gaussian pulse is defined in Eq. (7). This Gaussian pulse interacts with the $x$-polarized cavity mode and can selectively excite the UP or the LP when the frequency is set as $\omega = \omega_{LP}$ or $\omega = \omega_{UP}$. The coupling coefficient between the Gaussian pulse and the cavity mode was set as $Q_c = Q_{k,\lambda} = 0.1$ a.u.; see Eq. (3b) and associated text for more details on the definition of $Q_{k,\lambda}$.

During all these QM/MM simulations, an open boundary condition was used. Therefore,

the MM solvent molecules could potentially evaporate when the simulation time becomes very large. However, we checked that during our simulation timescale, i.e, tens of ps, no significant evaporation effect was observed, and each $Fe(CO)_5$ molecule was always surrounded by MM molecules. Hence, the use of the open boundary condition is valid for the current study.

When studying the Berry pseudorotation reaction, for each $Fe(CO)_5$ geometry, we defined the axial CO ligands as the ligand pair with the largest C−Fe−C angle. In Fig. 5, the number of barrier crossings along a 2 ps time span was computed as the sum of the crossings occurring over intervals of 0.5 ps for all of the independent molecular systems. For example, the number of barrier crossings at $t = 1.5$ ps was calculated as the sum of the total number of pseudorotation reactions occurring between $t = 0$ and $t = 0.5$ ps, between $t = 0.5$ and $1.0$ ps, and between $t = 1.0$ and $t = 1.5$ ps.

Additional QM simulations excluding the MM solvent molecules were also propagated to calculate the IR spectrum in Figs. 2a,b. These purely QM simulations were propagated in the same manner as the above QM/MM simulations.

## 2. Hopfield coefficients

We use the following phenomenological three-state model (with $\hbar = 1$) to describe the interaction between the cavity mode and the equatorial and axial vibration modes:

$$\hat{H}_{\text{model}} = \begin{pmatrix} \omega_{\text{c}} & g_{\text{e}} & g_{\text{a}} \\ g_{\text{e}} & \omega_{\text{e}} & 0 \\ g_{\text{a}} & 0 & \omega_{\text{a}} \end{pmatrix}. \tag{S2}$$

Here, $\omega_{\text{c}}$, $\omega_{\text{e}}$, and $\omega_{\text{a}}$ denote the frequencies of the cavity mode, the equatorial CO vibrational mode, and the axial CO vibrational mode, respectively; $g_{\text{e}}$ and $g_{\text{a}}$ denote the corresponding light-matter couplings. The polariton frequencies are defined as the eigen frequencies of

the three-state model. From the Hessian calculation of a single $Fe(CO)_5$ in vacuum at its equilibrium geometry, we obtain $\omega_e = 2010$ cm$^{-1}$, $\omega_a = 2025$ cm$^{-1}$, and $g_e/g_a = \sqrt{I_e/I_a} = 1.31$, where $I_e$ and $I_a$ denote the IR intensities of the equatorial and axial vibrational modes, respectively. Given $\omega_c = 2018$ cm$^{-1}$, we find that when $g_a = 69$ cm$^{-1}$, the calculated Rabi splitting matches the Rabi splitting in Fig. 2d well. The eigenvalues of our model are $\omega_{LP} = 1903$ cm$^{-1}$, $\omega_{MP} = 2019$ cm$^{-1}$, $\omega_{UP} = 2131$ cm$^{-1}$, and the three eigenvectors are

$$v_{LP} = \begin{pmatrix} -0.70 \\ 0.59 \\ 0.40 \end{pmatrix}, v_{MP} = \begin{pmatrix} -0.06 \\ -0.61 \\ 0.79 \end{pmatrix}, v_{UP} = \begin{pmatrix} 0.71 \\ 0.53 \\ 0.46 \end{pmatrix}. \tag{S3}$$

For each eigenstate, the weights of the photon or two vibrational modes, also known as the Hopfield coefficients, are

$$v_{LP}^2 = \begin{pmatrix} 0.49 \\ 0.35 \\ 0.16 \end{pmatrix}, v_{MP}^2 = \begin{pmatrix} 0.004 \\ 0.37 \\ 0.63 \end{pmatrix}, v_{UP}^2 = \begin{pmatrix} 0.50 \\ 0.28 \\ 0.21 \end{pmatrix}. \tag{S4}$$

Hence, for the UP, the equatorial weight is 0.28, the axial weight is 0.21, and their ratio is 1.33; for the LP, the equatorial weight is 0.35, the axial weight is 0.16, and their ratio is 2.2. The photonic weight in the MP is nearly zero, indicating a very small MP peak in the photonic signal.

# 3. Polariton dynamics for $Fe(CO)_5$ in vacuum

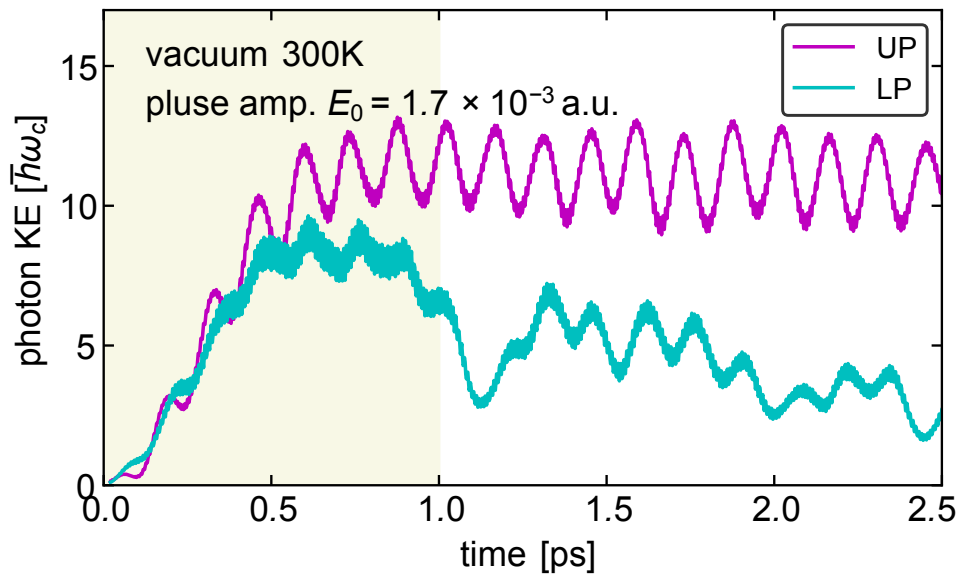

Figure S1: Kinetic energy dynamics of cavity photons under Gaussian pumping of the UP (magenta curve) or LP (cyan curve) in the absence of $n$-dodecane solvent molecules. All simulation conditions are the same as Fig. 3a except here the solvent molecules are not included (i.e., the simulation is performed for the solute in vacuum). Compared with Fig. 3a in the main text, excluding $n$-dodecane solvent molecules significantly slows down the UP dephasing but has less impact on the LP dephasing.

# 4. Benchmark of single-molecule reaction under VSC

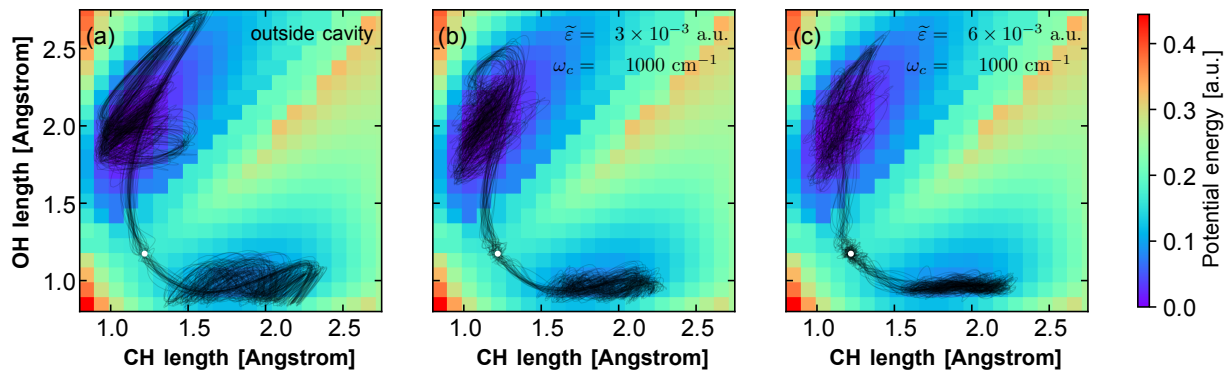

Figure S2: VSC effects on single-molecule isomerization reaction: $CH_2O \longleftrightarrow CHOH$. The molecule starts at the transition state (the white dot in each plot), and the cavity frequency is $\omega_c = 1000$ cm$^{-1}$. The molecule is (a) decoupled from the cavity or (b,c) coupled to the cavity with light-matter coupling strength (b) $\tilde{\varepsilon} = 3 \times 10^{-3}$ a.u. or (c) $\tilde{\varepsilon} = 6 \times 10^{-3}$ a.u. Each thin black line represents a molecular trajectory when the initial velocities are sampled from a Maxwell–Boltzmann distribution at 300 K, and 64 trajectories are sampled per panel. Under VSC, the molecular trajectory spends more time near the transition state, in agreement with a recent model study of the dynamical caging effect.[S9] Note that this study is used to benchmark the functionality of CavMD and should not be used to interpret recent VSC experiments on chemical catalysis in the collective regime. A tutorial on running this calculation is also available at `https://github.com/TaoELi/cavity-md-ipi`. Psi4[S10] is used to perform on-the-fly gradient evaluations at the level of Hartree–Fock theory with a 6-31G* basis set.



\end{document}